\documentclass[conference]{IEEEtran}
\IEEEoverridecommandlockouts

\usepackage[
    a4paper,
    left=13.2mm,
    right=13.2mm,
    bottom=42.3mm,
    top=16.7mm,
]{geometry}

\usepackage{lipsum}
\usepackage{multicol}
\usepackage{algorithm}
\usepackage{algpseudocode}
\usepackage{cite}
\usepackage{tabularx}
\usepackage{booktabs}
\usepackage{pdflscape}
\usepackage{mathtools}
\usepackage[square, numbers, sort]{natbib}
\usepackage{multirow}
\usepackage{comment}
\usepackage{amsmath,amssymb,amsfonts}
\usepackage{graphicx}
\usepackage{textcomp}
\usepackage{xcolor}
\usepackage{url}
\usepackage{float}
\usepackage{hyperref}
\usepackage{caption}
\usepackage{subcaption}
\usepackage{tikz}
\usetikzlibrary{arrows.meta, calc, positioning, shapes.geometric, fit, backgrounds}
\usepackage{forest}
\usepackage{graphicx} 
\usepackage{longtable}
\usepackage{ltablex} 
\usepackage{booktabs}
\usepackage{geometry}

\keepXColumns
\usepackage{xcolor}
\usepackage{xcolor}
\colorlet{myred}{red!80!black}
\colorlet{myblue}{blue!80!black}
\colorlet{mygreen}{green!60!black}
\colorlet{myorange}{orange!70!red!60!black}
\colorlet{mydarkred}{red!30!black}
\colorlet{mydarkblue}{blue!40!black}
\colorlet{mydarkgreen}{green!30!black}

\usepackage{amsmath} 
\usepackage{listofitems} 
\usepackage[outline]{contour} 
\contourlength{1.4pt}

\tikzset{
  >=latex, 
  node/.style={
    thick,
    circle,
    draw=myblue,
    minimum size=22,
    inner sep=0.5,
    outer sep=0.6,
    font=\small
  },
  node in/.style={node, fill=mygreen!25, draw=mygreen!70!black},
  node hidden/.style={node, fill=myorange!20, draw=myorange!70!black},
  node out/.style={node, fill=myred!20, draw=myred!70!black},
  connect/.style={thick,mydarkblue},
  connect arrow/.style={-{Latex[length=4,width=3.5]},thick,mydarkblue,shorten <=1,shorten >=1},
  box/.style={draw, rounded corners=2, inner sep=2pt},
}

\def\BibTeX{{\rm B\kern-.05em{\sc i\kern-.025em b}\kern-.08em
    T\kern-.1667em\lower.7ex\hbox{E}\kern-.125emX}}

\begin{document}

\title{Dueling Deep Q-Learning for Intrusion Detection}

\author{
\IEEEauthorblockN{Logan Luna\thanks{L. Luna is now with the School of Computer Science, College of Computing, Georgia Institute of Technology, Atlanta, GA 30332 USA (e-mail: lluna@gatech.edu).}\thanks{\copyright~2025 IEEE. Personal use of this material is permitted. Permission from IEEE must be obtained for all other uses, in any current or future media, including reprinting/republishing this material for advertising or promotional purposes, creating new collective works, for resale or redistribution to servers or lists, or reuse of any copyrighted component of this work in other works. Published version: DOI 10.1109/SOUTHEASTCON56624.2025.10971436}}
\IEEEauthorblockA{
\textit{Department of Electrical Engineering and Computer Science} \\
\textit{Embry-Riddle Aeronautical University}\\
Daytona Beach, USA}
\and
\IEEEauthorblockN{Matthew P. Berkowitz}
\IEEEauthorblockA{
\textit{Department of Electrical Engineering and Computer Science} \\
\textit{Embry-Riddle Aeronautical University}\\
Daytona Beach, USA \\
berkowim@my.erau.edu}
\and
\IEEEauthorblockN{Laxima Niure Kandel}
\IEEEauthorblockA{
\textit{Department of Electrical Engineering and Computer Science} \\
\textit{Embry-Riddle Aeronautical University}\\
Daytona Beach, USA \\
niurekal@erau.edu}
\and
\IEEEauthorblockN{Sirio Jansen-Sánchez}
\IEEEauthorblockA{
\textit{Department of Electrical Engineering and Computer Science} \\
\textit{Embry-Riddle Aeronautical University}\\
Daytona Beach, USA \\
jansenss@my.erau.edu}
}



\maketitle

\begin{abstract}

Intrusion detection systems (IDS) and automated systems for detecting and reporting cyber threats, are commonly handled via supervised machine learning methods. Though effective, these models struggle to effectively adapt to new attack types. This study proposes a novel approach by employing a reward-based, dueling Q-learning model for IDS, achieving an average accuracy of 99.68\% across multiple attack classes. The proposed model has a dueling network architecture which separates its predictions into value and advantage streams. This has the benefit of improving learning efficiency and stability. The model was trained on the CIC-IDS2018, a benchmark dataset based on real-world intrusion detection scenarios, having multiple attack classes such as DDoS, botnets, and brute-force attacks. Furthermore, Explainable AI  (XAI), specifically SHAP (SHapley Additive exPlanations), was also integrated into the training and evaluation process to provide interpretability into the model's predictions.

\end{abstract}


\section{Introduction}

\subsection{The Growing Complexity of Network Threats}\label{A}

Modern network environments face a growing range of sophisticated and diverse cyber threats. Traditional signature-based detection methods rely on prior knowledge of attack patterns, however, this limits their ability to recognize novel or evolving threats \cite{khraisat2019survey}. Similar to this, approaches using anomaly detection are capable of detecting novel or evolving threats, but face issues with high false-positive rates \cite{garcia2009anomaly}. This issue is further augmented in dynamic and diverse environments \cite{garcia2009anomaly}. Traditional intrusion detection systems (IDS) have been noted for inefficiency in handling polymorphic attacks and zero-day exploits \cite{sommer2010}. Furthermore, Chandola et al. \cite{chandola2009} outlines the limitations of anomaly detection in dynamic environments, highlighting the critical need for models that minimize false positives while maintaining sensitivity.

Foundational research on network security highlights these challenges, emphasizing the criticality of developing systems that can adapt to diverse and evolving threats in real-time scenarios \cite{estevez2004techniques}. Industry reports further illuminate this need, such as the Verizon Data Breach Investigations Report \cite{verizon2022}, which discusses the evolving landscape and inadequacies of current systems.These  limitations underscore the urgent need for an adaptive solutions that can effectively respond to new attack vectors.  
\vspace{-0.3cm}
\subsection{The Promise of Reinforcement Learning}\label{B}

Reinforcement Learning (RL) offers a promising alternative to current implementations to address the outlined limitations in ~\ref{A}. Since the model learns from a reward structure rather than being based on purely labeled data, it optimizes its policy around the influence of its action on the environment rather than just classifying it correctly \cite{sutton1998reinforcement}. As a result, this makes RL more adaptable, making it suitable for environments where attack types constantly evolve. Furthermore, the customizable reward structure allows the model to be further adjusted towards a specific desired policy \cite{sutton1998reinforcement}, overall offering an effective alternative for IDSs. 
\subsection{Our Contributions:}
\begin{itemize}
    \item \textbf{Dueling Network Architecture}: 
    By employing a dueling architecture which separates the value stream to estimate the value of the current state \( V(\mathbf{s}) \), independent of actions, and the advantage streams to estimate the advantage \( A(\mathbf{s}, a) \) of each action, representing the relative importance of actions in a given state, and combining both streams to compute the final Q-values, overall enhancing stability and convergence during training, the proposed model has improved learning stability and efficiency, 
    
    \item \textbf{Training on Real-World Data}: The model was trained on the CIC-IDS2018, a benchmark dataset representing real-world intrusion detection scenarios, spanning multiple attack types such as DDoS, botnets, and brute-force attacks. Altogether the study utilized 2,177,804 samples. 
    
    \item \textbf{Explainability with XAI}: To improve interpretability, SHAP (SHapley Additive exPlanations) was utilized to provide insight into the key features influencing the model’s predictions and illustrating their impact on the model's decision-making process. 
  
    \item \textbf{High Performance Across Attack Classes}: The proposed model achieved an average accuracy of 99.68\% across multiple attack types. This outperforms previously implemented RL-based methods for IDSs \cite{alavizadeh2022deep} \cite{Singh2023}. 
    
\end{itemize}

\section{Background and Related Work}

\tikzset{%
   neuron missing/.style={
    draw=none, 
    scale=2,
    text height=0.333cm,
    execute at begin node=\color{black}$\vdots$
  },
}

\newcommand{\DrawNeuronalNetwork}[2][]{
\xdef\Xmax{0}
\foreach \Layer/\X/\Col/\Miss/\Lab/\Count/\Content [count=\Y] in {#2}
{\pgfmathsetmacro{\Xmax}{max(\X,\Xmax)}
\xdef\Xmax{\Xmax}
 \xdef\Ymax{\Y}
}
\foreach \Layer/\X/\Col/\Miss/\Lab/\Count/\Content [count=\Y] in {#2}
{\node[anchor=south] at ({2*\Y},{\Xmax/2+0.1}) {\Layer};
 \foreach \m in {1,...,\X}
 {
  \ifnum\m=\Miss
   \node [neuron missing] (neuron-\Y-\m) at ({2*\Y},{\X/2-\m}) {};
  \else
   \node [circle,fill=\Col!50,minimum size=1cm] (neuron-\Y-\m) at 
  ({2*\Y},{\X/2-\m}) {\Content};
   \ifnum\Y=1
   \else
    \pgfmathtruncatemacro{\LastY}{\Y-1}
    \foreach \Z in {1,...,\LastX}
    {
     \ifnum\Z=\LastMiss
     \else
      \draw[->] (neuron-\LastY-\Z) -- (neuron-\Y-\m);
     \fi
    }
   \fi
  \fi
 }
 \xdef\LastMiss{\Miss}
 \xdef\LastX{\X}
}
}

\subsection{Reinforcement Learning}
As described in ~\ref{B}, RL is a paradigm within machine learning where an agent learns to make decisions by interacting with an environment to maximize cumulative rewards. 



\begin{figure}[ht!]
    \centering
    \begin{tikzpicture}[scale=0.45, transform shape, x=1.5cm, y=1.5cm, >=stealth, font=\sffamily, nodes={align=center}]
        \begin{scope}[local bounding box=T]
            \path node[draw, minimum width=6em, minimum height=4em] (state) {State};
            \begin{scope}[local bounding box=NN]
                \DrawNeuronalNetwork{Input Layer/5/green/4///,
                  Hidden Layer/5/blue/4//11/,
                  Output Layer/4/red/3//11/}
            \end{scope}
            \path (NN.south) node[below]{Estimated parameter\\ $\theta$};
            \path(NN.east) -- node[above]{Policy\\ $\Pi(\theta,a)$}++ (4em,0);
        \end{scope} 
        \node[fit=(T), label={[anchor=north west]north west:Agent}, inner sep=1em, draw] (TF) {};
        \node[below=3em of TF, draw, inner sep=1em] (Env) {Environment};
        \draw[<-] (TF.200) -- ++(-1em,0) |- (Env.160) node[pos=0.45,right]{$r_t$};
        \draw[<-] (TF.180) -- ++(-2em,0) |- (Env.180) node[pos=0.45,left]{$s_t$};
        \draw[->] (NN.east) -- ++(7em,0) node[right]{$a_t$} |- (Env);
    \end{tikzpicture}
    \caption{RL architecture visualization with input, hidden, and output layers connected to an environment through agent interaction \cite{sutton1998reinforcement}.}
    \label{fig:neural_network}
\end{figure}
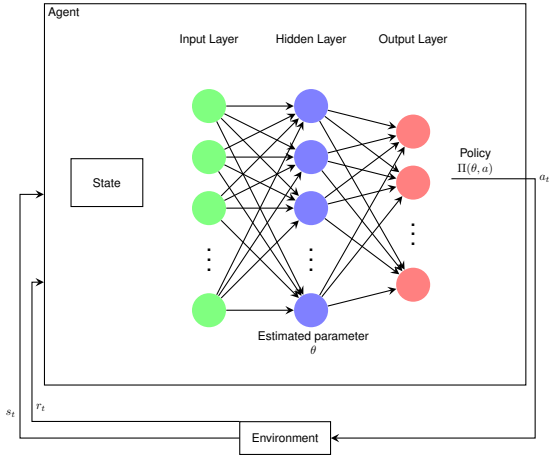

\subsubsection{Environment, States, and Agent}
In RL, the \textbf{environment} represents the external system with which the agent interacts. It provides \textbf{states} ($s_t$) that describe the current situation and returns \textbf{rewards} ($r_t$) based on the agent's actions ($a_t$). The \textbf{agent} is the decision-maker that selects actions to maximize the expected cumulative reward. The interaction loop involves the agent perceiving the state, choosing an action, and receiving feedback from the environment, as depicted in Fig. \ref{fig:neural_network}.

\subsubsection{Actions and Rewards}
Actions are the possible moves or decisions the agent can make in each state. The choice of actions influences the next state and the rewards received which guide the agent towards desirable behaviors. The agent's objective is to learn a policy $\Pi$ that maps states to actions, maximizing the long-term reward.


Previous literature exploring the application and advantages of RL in IDSs, such as those by Alavizadeh et al. \cite{alavizadeh2022deep} and Singh et al. \cite{Singh2023} present several limitations.
For instance, Alavizadeh et al. \cite{alavizadeh2022deep} achieved moderate accuracy (e.g., ~88\%) in multi-class classifications. In contrast, traditional IDS models often demonstrate high accuracy, as demonstrated by Songma et al. \cite{optimizedIDS2023}, who achieved high accuracy of  ~99\% using random forest classifier. While Singh et al. \cite{Singh2023} achieved higher-performance using RL, their approach primarily focused on specific situations such as agent versus agent interactions, rather than addressing more common attack types such as DDOS or Brute force.

\subsection{Q-Learning and Deep Q-Networks (DQN)}
\subsubsection{Q-Learning}
Q-Learning is a foundational model-free RL algorithm aimed at learning the optimal action-value function \( Q(s, a) \), which estimates the expected return of taking action \( a \) in state \( s \). The update rule is given by:
\begin{equation}
    Q(s, a) \leftarrow Q(s, a) + \alpha \left[ r + \gamma \max_{a'} Q(s', a') - Q(s, a) \right]
\end{equation}
where \( \alpha \) is the learning rate and \( \gamma \) is the discount factor. Q-Learning utilizes a Q-table to store and update these values, but it becomes infeasible in high-dimensional spaces.

\subsubsection{Deep Q-Networks (DQN)}
Deep Q-Networks address the scalability issue of Q-Learning by employing deep neural networks to approximate the \( Q(s, a) \) function. Introduced by Mnih et al. \cite{mnih2015dqn}, DQNs can handle large state-action spaces by generalizing across similar states. This approach offers significant benefits for applications like intrusion detection by effectively modeling complex, non-linear relationships and enhancing scalability.

\subsubsection{Dueling Q-Networks}
Dueling Q-Networks extend DQNs by decomposing the \( Q(s, a) \) function into separate estimators for the state value \( V(s) \) and the advantage \( A(s, a) \):
\begin{equation}
    Q(s, a) = V(s) + \left( A(s, a) - \frac{1}{|\mathcal{A}|} \sum_{a'} A(s, a') \right)
\end{equation}
where \( |\mathcal{A}| \) is the number of possible actions. This formulation ensures that the value and advantage streams contribute appropriately to the final Q-values. This architecture, proposed by Wang et al. \cite{wang2016dueling}, improves learning stability and efficiency by allowing the network to focus on the most relevant actions. Benefits include enhanced generalization, reduced training instability, and faster convergence, particularly in environments with subtle action differences.

\subsection{Dataset (CIC-IDS2018) and Experimental Setup}

This study utilized the CIC-IDS 2018 dataset, created by Sharafaldin et al. \cite{CIC2018}. The dataset has a wide set of labeled network traffic data, containing various benign and malicious data. The dataset is created to represent a real-world environment, making it suitable for training and evaluating intrusion detection models. The distribution of data in this dataset is shown in Table \ref{tab:dataset}. This dataset has been used extensively in the development of IDS. Such instances include \cite{optimizedIDS2023}, which utilized a Random Forest classifier to achieve a 99\% accuracy. Additionally, \cite{treebasedIDS2023} employed ensemble methods using Decision Tree and XGBoost classifiers, reaching a 98.36\% accuracy. Lastly, HAST-IDS employed a hierarchical spatial-temporal feature extraction to achieve 92\% accuracy. These high results are consistent with general IDS research, where ML models commonly reach high performance due to the quality and variety of features in these datasets.



\renewcommand{\arraystretch}{1.3} 
\setlength{\tabcolsep}{5pt} 
\small

\begin{table}[htbp]
\centering
\caption{Dataset Attack Type Distribution}
\label{tab:dataset}
\begin{tabular}{p{2cm}p{3cm}p{2cm}}
\toprule
\textbf{Attack Type} & \textbf{Definition} & \textbf{Samples} \\
\midrule
Benign        & Normal activities. & 1,935,399 \\
DDoS Attack   & Distributed DoS involving multiple sources. & 155,191 \\
DoS Attack    & Attack aimed at disrupting access to a service. & 39,314 \\
Botnet        & Use of compromised devices to perform malicious activities. & 28,907 \\
Brute-force   & Repeated attempts to gain unauthorized access. & 18,820 \\
Web Attack    & Attacks targeting web servers, such as injections. & 173 \\
\bottomrule
\end{tabular}
\end{table}

\section{Methodology}

\subsection{Dueling Q-Network Framework}
This study proposes a Dueling Q-Network (DQN) Framework for intrusion detection. This approach aims to improve scalability and accuracy by employing a DQN. The Agent uses value and advantage streams to provide detailed classification for specific attacks, and has the parameters outlined in Table \ref{tab:model_parameters}.

\renewcommand{\arraystretch}{1.3} 
\setlength{\tabcolsep}{5pt} 
\small
\begin{table}[htbp]
\centering
\caption{Dueling DQN Model Parameters}
\label{tab:model_parameters}
\begin{tabular}{p{4cm}p{4cm}}
\toprule
\textbf{Parameter} & \textbf{Value} \\
\midrule
Model Type          & Dueling DQN \\
Hidden Layers      & [128, 64] \\
Batch Size         & 128 \\
Learning Rate      & 0.001 \\
Gamma (Discount)   & 0.99 \\
Epsilon Start      & 1.0 \\
Epsilon End        & 0.1 \\
Epsilon Decay      & 0.999 \\
Memory Size        & 10,000 experiences \\
Target Update Freq & Every 1000 steps \\
Episode Count      & 200 \\
Device            & CUDA GPU Nvidia 3060 \\
Optimizer         & Adam \\
Loss Function     & MSE \\
\bottomrule
\end{tabular}
\end{table}

\subsubsection{Dueling Deep Q-Network}
This agent identifies specific attack types, providing higher accuracy than traditional models but also has increased computational requirements.

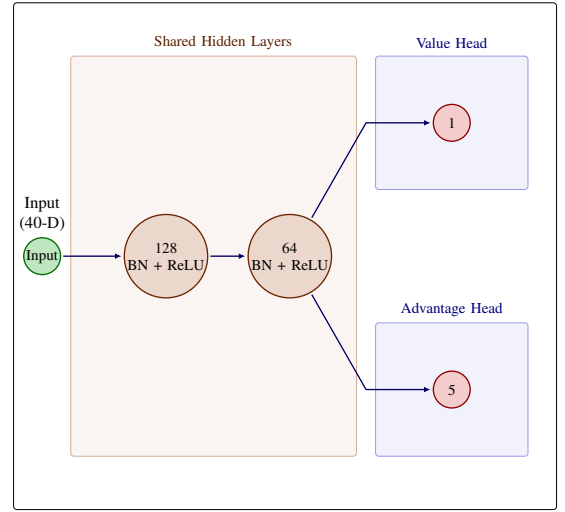
\begin{figure}[htbp]
    \centering

    \begin{subfigure}{0.4\textwidth}
        \centering
        \resizebox{\textwidth}{!}{
            \begin{tikzpicture}[x=2.0cm,y=2.8cm, every node/.style={transform shape, align=center}]
                \draw[box] (-0.3,-0.4) rectangle (5.4,3.4);

                \draw[myorange!40,fill=myorange!05,rounded corners=2]
                (0.3, -0.0) rectangle (3.3, 3.0);
                \node[myorange!60!black] at (1.9,3.1) {\small Shared Hidden Layers};

                \draw[myblue!40,fill=myblue!05,rounded corners=2]
                (3.5,-0.0) rectangle (5.1,1.0);
                \draw[myblue!40,fill=myblue!05,rounded corners=2]
                (3.5, 2.0) rectangle (5.1,3.0);
                \node[myblue!60!black] at (4.3,3.1) {\small Value Head};
                \node[myblue!60!black] at (4.3,1.1) {\small Advantage Head};

                \node[node in, label={above:\shortstack{Input\\(40-D)}}] (input) at (0,1.5) {Input};

                \node[node hidden] (h1) at (1.3, 1.5) {128 \\ BN + ReLU};

                \node[node hidden] (h2) at (2.6, 1.5) {64 \\ BN + ReLU};

                \node[node out] (val) at (4.3, 2.5) {1};

                \node[node out] (adv) at (4.3, 0.5) {5};

                \draw[connect arrow] (input) -- (h1);
                \draw[connect arrow] (h1) -- (h2);
                \draw[connect arrow] (h2) -- ++(0.8,1.0) -- (val);
                \draw[connect arrow] (h2) -- ++(0.8,-1.0) -- (adv);
            \end{tikzpicture}
        }
    \end{subfigure}

    \caption{Architecture of Dueling DQN (shared layers feeding into value (1-dim) and advantage (5-dim) outputs) Agent}
        \label{fig:low-level-agent}
\end{figure}

\begin{figure}[htbp]
\centering

\begin{tikzpicture}[scale=0.70, x=2.7cm,y=1.6cm]
  \def\NI{5}   
  \def\NO{4}   
  \def\yshift{0.4} 
  
  \foreach \i [evaluate={\c=int(\i==\NI);
                         \y=\NI/2-\i-\c*\yshift; 
                         \index=(\i<\NI?int(\i):"n");}]
              in {1,...,\NI}{ 
    \node[node in] (NI-\i) at (0,\y) {$a_{\index}^{(0)}$};
  }
  
  \foreach \i [evaluate={\c=int(\i==\NO);
                         \y=\NO/2-\i-\c*\yshift; 
                         \index=(\i<\NO?int(\i):"m");}]
              in {\NO,...,1}{ 
    \ifnum\i=1
      \node[node hidden] (NO-\i) at (1,\y) {$a_{\index}^{(1)}$};
      \foreach \j [evaluate={\indexA=(\j<\NI?int(\j):"n");}] in {1,...,\NI}{ 
        \draw[connect,white,line width=1.2] (NI-\j) -- (NO-\i);
        \draw[connect] (NI-\j) -- (NO-\i)
          node[pos=0.5,above,scale=0.8,sloped]
            {\contour{white}{$w_{1,\indexA}$}};
      }
    \else
      \node[node,blue!20!black!80,draw=myblue!20,fill=myblue!5]
        (NO-\i) at (1,\y) {$a_{\index}^{(1)}$};
      \foreach \j in {1,...,\NI}{ 
        \draw[connect,myblue!20] (NI-\j) -- (NO-\i);
      }
    \fi
  }
  
  \path (NI-\NI) --++ (0,1+\yshift) node[midway,scale=1.2] {$\vdots$};
  \path (NO-\NO) --++ (0,1+\yshift) node[midway,scale=1.2] {$\vdots$};

  \node[right=0cm of NO-2, align=left, text width=6cm] {
    \[
    a_{1}^{(1)} = \sigma\left( \sum_{i=1}^{n} w_{1,i} a_{i}^{(0)} + b_{1}^{(0)} \right)
    \]
    \[
    \mathbf{a}^{(1)} = \sigma\left( \mathbf{W}^{(0)} \mathbf{a}^{(0)} + \mathbf{b}^{(0)} \right)
    \]
  };

\end{tikzpicture}
\caption{A single layer’s forward pass: input vector $\mathbf{a}^{(0)}$ transforms into $\mathbf{a}^{(1)}$ via weights $w_{j,i}$, biases $b_j$, and an activation $\sigma(\cdot)$.}
\label{fig:activation}
\end{figure}
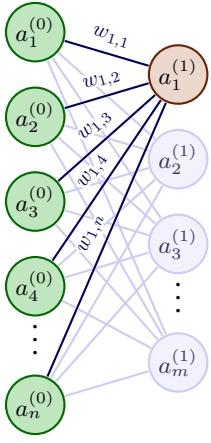

This agent employs a Dueling Deep Q-Network architecture for multiclass classifications. This chosen architecture enables the agent to further evaluate actions by decomposing the Q-value into a value stream to estimate the state value \( V(s) \), and an advantage stream to estimate the action advantage \( A(s, a) \). By separating these streams, the model can learn the value of states independent of the influence of actions. This has the effect of improving the learning efficiency of the model along with it's stability. A visualization of the dueling DQN is given in Fig. \ref{fig:low-level-agent}, with the shared hidden layers being composed of nodes as shown in Fig. \ref{fig:activation}. Furthermore, by using a deep network, the hidden layers process the state \( \mathbf{s} \) to provide a unified feature representation. A sample of a layer pass in this network is given by Fig. \ref{fig:activation}.

\textbf{Dueling Q Structure:}
\begin{itemize}
    \item \textbf{Value Stream}: Estimates the value of the current state \( V(\mathbf{s}) \), independent of actions.
    \item \textbf{Advantage Stream}: Estimates the advantage \( A(\mathbf{s}, a) \) of each action, representing the relative importance of actions in a given state.
    \item \textbf{Q-Value Computation}: Combines both streams to compute the final Q-values, enhancing stability and convergence during training.
\end{itemize}    

\begin{algorithm}
\caption{Deep Q-Network Architecture}
\begin{algorithmic}[1]
\State \textbf{Input}: Feature vector of size $40$
\State \textbf{Output}: Value and Advantage estimates
\Statex
\State \textbf{Shared Layers:}
\State $x \gets \text{Linear}(40, 128)$
\State $x \gets \text{BatchNorm1d}(128)$
\State $x \gets \text{ReLU}(x)$
\State $x \gets \text{Linear}(128, 64)$
\State $x \gets \text{BatchNorm1d}(64)$
\State $x \gets \text{ReLU}(x)$
\Statex
\State \textbf{Value Stream:}
\State $V \gets \text{Linear}(64, 1)$
\Statex
\State \textbf{Advantage Stream:}
\State $A \gets \text{Linear}(64, 5)$
\Statex
\State \textbf{Q-Value:}
\State $Q(s, a) = V + \left(A - \frac{1}{|\mathcal{A}|} \sum_{a'} A(s, a') \right)$
\end{algorithmic}
\end{algorithm}

\subsection{Environment Design}

To evaluate the proposed framework, a custom environment, \texttt{NetworkClassificationEnv}, was designed using OpenAI Gym. This environment facilitates the interaction of the outlined agents with network traffic data for both classification tasks and reward-based training. The most influential features utilized in this environment are outlined in Table \ref{tab:feature-attributes}, however in total there are over 80 features. 

\texttt{NetworkClassificationEnv} processes network traffic sequentially, treating each data point as an individual network flow based on its timestamp. This setup supports the DQN for making reward-based decisions, aligning with the agent structure.

\renewcommand{\arraystretch}{1.3} 
\setlength{\tabcolsep}{5pt} 
\small

\begin{table}[htbp]
\centering
\caption{Feature Attributes in NetworkClassificationEnv}
\label{tab:feature-attributes}
\begin{tabular}{p{2cm}p{3cm}p{1.5cm}}
\toprule
\textbf{Feature Name} & \textbf{Description} & \textbf{Type} \\
\midrule
\texttt{Flow Duration}  & Total duration of the flow & Continuous \\
\texttt{Flow Byts/s}    & Byte rate of the flow & Continuous \\
\texttt{Flow Pkts/s}    & Packet rate of the flow & Continuous \\
\texttt{Pkt Len Mean}   & Mean packet length & Continuous \\
\texttt{Pkt Len Std}    & Standard deviation of packet length & Continuous \\
\texttt{IAT Mean}       & Mean inter-arrival time & Continuous \\
\texttt{SYN Flag Cnt}   & Count of SYN flags & Integer \\
\bottomrule
\end{tabular}
\end{table}

\textbf{Key Features}:  
\begin{itemize}
    \item \textbf{Batch Processing}: Handles network flows in configurable batch sizes, enhancing training efficiency through parallel processing.
    \item \textbf{Reward Calculation}: The reward function is structured as follows:
    
    \[
    \small
    r_t = 
    \begin{cases} 
    +1 \cdot S_l \cdot C_a + \min(0.5 \cdot \log(\text{streak}), 2.0) & \text{if } a_t = l_t, \\
    -1 \cdot S_l \cdot C_a & \text{if } a_t \neq l_t, \\
    0 & \text{if no action}.
    \end{cases}
    \]
    
    \text{where:}
    \begin{align*}
    S_l & = \text{Severity weight based on the true label } l_t, \\
    C_a & = 0.5 + \frac{\text{confidence}}{2}, \\
    \text{streak} & = \text{number of consecutive correct predictions}.
    \end{align*}
\end{itemize}

\section{Results}

\begin{table*}[h!]
    \centering
    \caption{Comparison of Related Work in Network Intrusion Detection}
    \label{tab:comparison}
    \resizebox{0.97\textwidth}{!}{ 
    \begin{tabular}{|l|l|l|l|l|}
        \hline
        \textbf{Study} & \textbf{Dataset} & \textbf{Methodology} & \textbf{Key Limitations} & \textbf{Performance} \\
        \hline
        Our Study & CIC-IDS2018 & Reward-Based Deep Q-Learning & 
        \begin{tabular}[c]{@{}l@{}}- Preliminary results without environment adaptation\end{tabular} & 
        \begin{tabular}[c]{@{}l@{}}
        99.68\% \end{tabular} \\ 
        \hline
        Alavizadeh et al.~\cite{alavizadeh2022deep} & NSL-KDD & Deep Q-Learning & 
        \begin{tabular}[c]{@{}l@{}}- Q-learning used as a supervised classifier\\ 
        - Smaller sample size\end{tabular} & 88\% \\
        \hline
        Optimized IDS~\cite{optimizedIDS2023} & CIC-IDS2018 & Random Forest & 
        \begin{tabular}[c]{@{}l@{}}- No temporal data handling\\ 
        - Focused on feature selection and classification\end{tabular} & 99\% \\
        \hline
        Tree-Based IDS~\cite{treebasedIDS2023} & CIC-IDS2018 & Ensemble (Decision Tree, XGBoost) & 
        \begin{tabular}[c]{@{}l@{}}- High computational cost for training\\ 
        - Limited generalizability to dynamic environments\end{tabular} & 98.36\% \\
        \hline
        HAST-IDS~\cite{Zhao2018} & Custom IIoT Dataset & Hierarchical Spatial-Temporal Features & 
        \begin{tabular}[c]{@{}l@{}}- Focused on feature extraction\\ 
        - Limited scalability for real-time applications\end{tabular} & 92\% \\
        \hline
    \end{tabular}
    }
\end{table*}

\subsection{Model Performance}
\begin{figure}[htbp]
    \centering
    \includegraphics[width=0.40\textwidth]{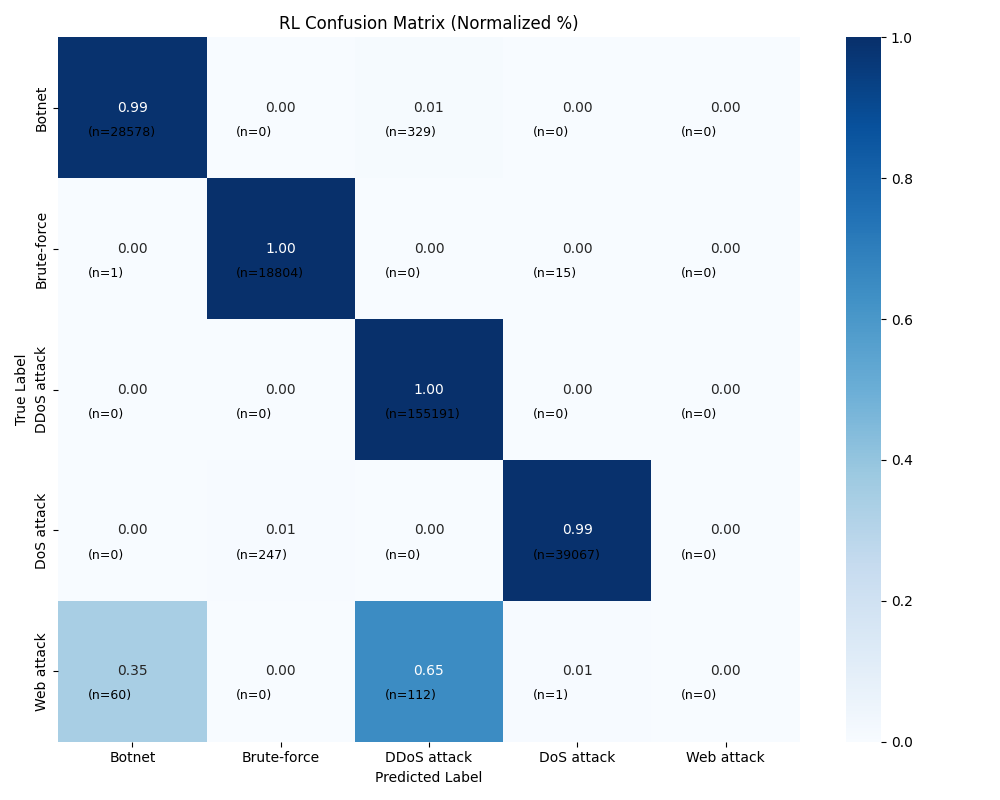}
    \caption{Confusion Matrix for DQN Agent}
    \label{fig:rl_confusion_matrix}
\end{figure}

Our results demonstrate that proposed DQN Agent achieves an average accuracy of 99.68\%, successfully classifying various attack types present in network traffic as shown in Table \ref{tab:multi_classification} and confusion matrix (see Fig.~\ref{fig:rl_confusion_matrix} ). It's important to note that, the \textit{Web attack} was excluded from analysis due to its limited sample size of just 173 samples. While web attacks are rare, there are many different approaches that were considered to be taken, such as weighting the reward function higher, or even generating mock data to help train the model. The problem with this is that mock data can lead to certain repeated features in the data being used to classify the attack instead of what features should actually be used.

\begin{table}[H]
    \centering
    \caption{Classification Report for DQN Agent}
    \label{tab:multi_classification}
    \begin{tabular}{lcccc}
        \hline
        \textbf{Class} & \textbf{Precision} & \textbf{Recall} & \textbf{F1-Score} & \textbf{Samples} \\
        \hline
        Benign & 0.9998 & 0.99996 & 0.99987 & 1,935,399 \\
        Botnet & 0.9979 & 0.9886 & 0.9932 & 28,907 \\
        Brute-force & 0.9870 & 0.9991 & 0.9931 & 18,820 \\
        DDoS attack & 0.9972 & 1.0000 & 0.9986 & 155,191 \\
        DoS attack & 0.9996 & 0.9937 & 0.9966 & 39,314 \\
        Web attack & 0.0000 & 0.0000 & 0.0000 & 173 \\
        \hline
        \textbf{Accuracy} & \multicolumn{4}{c}{0.99684 (Weighted Avg)} \\
        \hline
    \end{tabular}
\end{table}

 The proposed hierarchical Q-learning framework has displayed notable improvements over similar previous studies that sought to implement RL for IDSs as shown in Table \ref{tab:comparison}. While Alavizadeh et al.~\cite{alavizadeh2022deep} achieved an accuracy of 88\%, our DQN attains 99.68\% accuracy in multi-class classifications. We predict that the enhanced performance over previous studies such as Alavizadeh et al. \cite{alavizadeh2022deep} is likely due to the incorporation of a Dueling DQN architecture along with more samples being used to model the training environment. Altogether, Alavizadeh et al. used 219,980 data samples to train their model, whereas our study utilized 2,177,804 samples. Additionally, our architecture used the more complex dueling structure, providing the model with value and advantage streams to improve performance. Overall, these changes have the advantage of enabling the model to make better-informed decisions to detect evolving attack vectors.
\subsection{Explainability Results}

\begin{figure}[htbp]
    \centering

    \begin{subfigure}{0.35\textwidth}
        \centering
        \includegraphics[width=\linewidth]{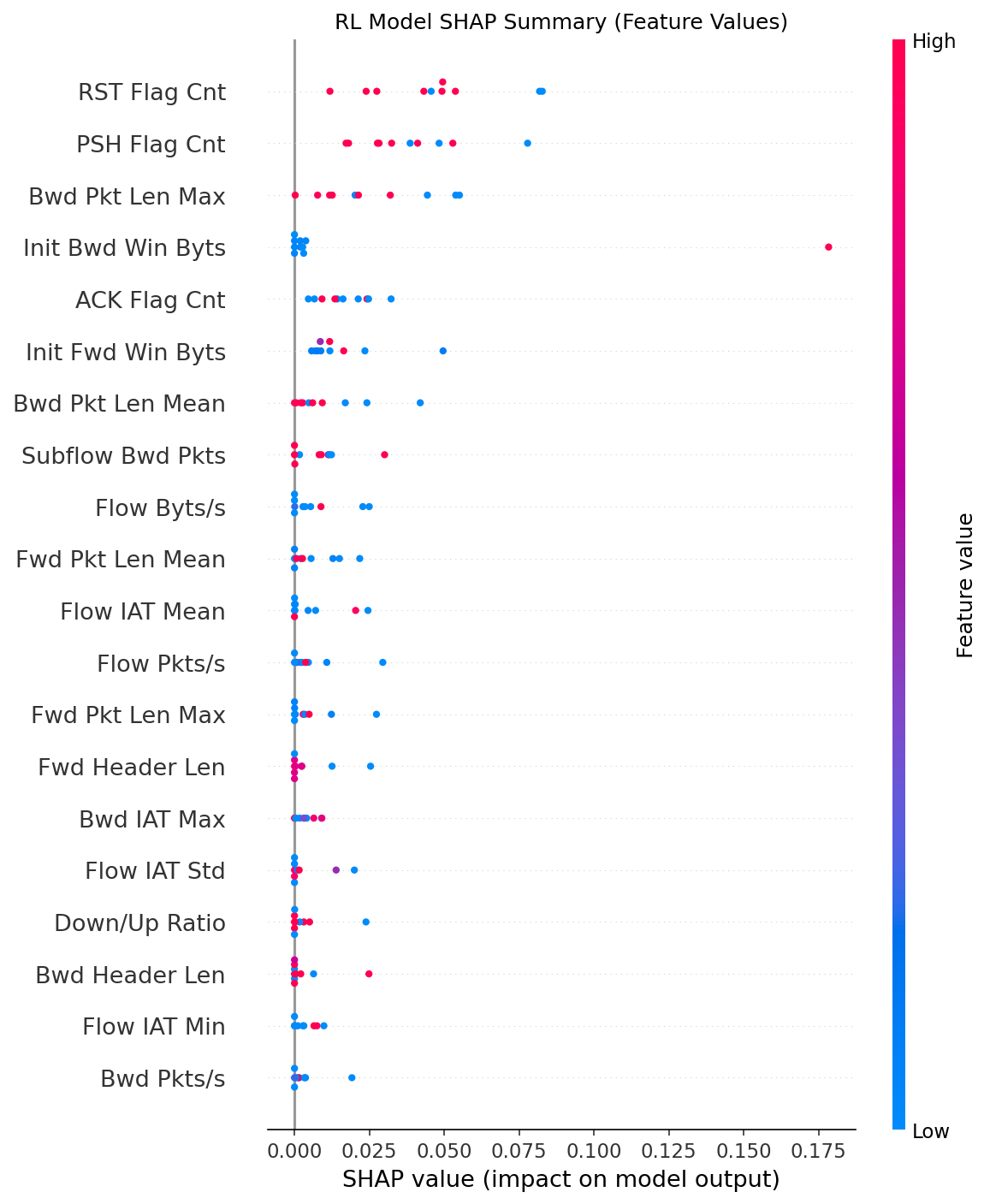}
        \caption{RL SHAP Beeswarm}
        \label{fig:rl_shap_beeswarm}
    \end{subfigure}
    \hfill
    \begin{subfigure}{0.35\textwidth}
        \centering
        \includegraphics[width=\linewidth]{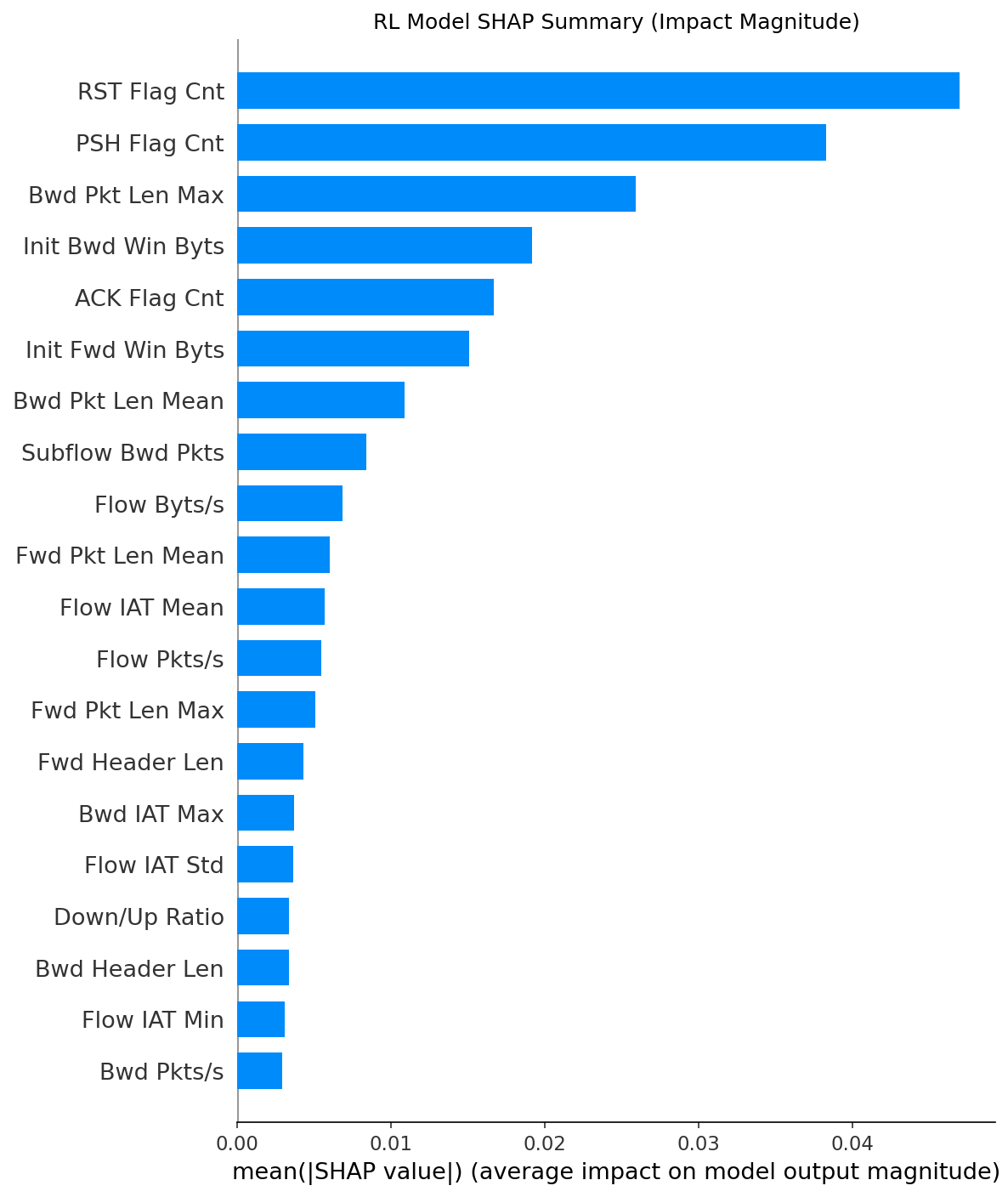}
        \caption{RL SHAP Summary}
        \label{fig:rl_shap_summary}
    \end{subfigure}
    \caption{Explainability resutls for DQN Decision Making (a) RL SHAP Beeswarm and (b) RL Summary Plots}
    \label{fig:shap_all_plots}
\end{figure}

Explainable AI (XAI), specifically SHAP (SHapley Additive exPlanations), were implemented into this pipeline for further analysis of the model's predictions and what factors influenced the decision-making as shown in Fig. \ref{fig:shap_all_plots}.\cite{beechey2023explaining}. 


\textbf{Fig. (\ref{fig:rl_shap_beeswarm}) SHAP Beeswarm Plot: }This plot displays how the features utilized influence the model’s decision-making. The X-axis is representative of the impact on the model's decision. The Y-axis ranks features by their overall importance. Each dot represents a data point, with color indicating the feature value (red = high, blue = low) \cite{beechey2023explaining}. An example of this is a a high (red) "Flow Pkts/s" value may indicate a DDoS attack, whereas a low (blue) value suggests normal traffic.

\textbf{Fig. (\ref{fig:rl_shap_summary}) SHAP Summary Plot: } This bar chart visualizes the influence of each feature on the model's decisions. This plot does not display the positive or negative impact of features as the Beeswarm plot does, instead only showing the magnitude of their contributions.

\subsubsection{QLearning SHAP Findings}
\begin{itemize}
    \item \textbf{RST Flag Cnt}: Highly impactful in identifying attacks like DoS or scanning, indicating abnormal termination of connections. This flag measures the number of TCP reset (RST) flags in a session. This indicates abrupt connection terminations, often seen in DoS attacks and network scanning to overwhelm a system. 

    \item \textbf{PSH Flag Cnt}: Reflects urgency in data transmission, often linked to buffer overflow or data exfiltration attempts. Such attacks that this can indicate are brute-force attacks where an attacker will attempt rapid authentication attempts, or botnet attacks where infected hosts rapidly send data to a command-and-control server.
    
    \item \textbf{Bwd Pkt Len Max}: Captures the largest backward packet being sent from server-to-client, often signifies exfiltration activities. This often indicates a botnet operations where stolen data is transferred from the compromised system.

    \item \textbf{Init Bwd Win Byts}: 
    Represents the initial window size received in the backward flow. Abnormal values are often associated with SYN flood DoS attacks, where malicious traffic will exploit handshake mechanisms to overwhelm server resources.

    \item \textbf{ACK Flag Cnt}: This flag measures acknowledgment packets and helps to differentiate benign flows from brute-force attacks based on handshake patterns. An irregular pattern can indicate frequent failed login attempts create unusual handshake behaviors, such as brute-force.

    \item \textbf{Flow Byts/s and Flow Pkts/s}: Measures the rate of packet traffic, elevated rates correlate with high-speed attacks like DDoS, indicating a spike in traffic to overwhelm a system.
\end{itemize}

\section{Limitations and Future Work}

Although the current framework demonstrates success in its initial implementation, there are several existing limitations that are to be addressed in future research.

\begin{itemize}

    \item \textbf{Limited Classification Granularity:} \\
    The current model performs only high-level classification of attack types, such as identifying an attack solely as DoS rather than DoS-Hulk or DoS-GoldenEye. This is a common approach, being done as well in Alavizadeh et al. \cite{alavizadeh2022deep}, however, we intend to fine-grained attack differentiation in the future by using advanced feature extraction techniques and hierarchical classification strategies.

 
    \item \textbf{Deployment in Real-World Scenarios:} \\
    This model was trained and evaluated in the same environment it was rained in. This does not fully capture its performance in a more complex and variable environment. 
    
    \item \textbf{Simulated Enterprise Environment:} 
    \\
    To further test the model, a simulated enterprise network environment will be set up with simulated normal net traffic. With this, the model can be tested with actual attacks. The model will act as an IDS and take the appropriate actions for each attack defending the environment.
    
\end{itemize}

\section{Conclusions}

In this work, we present a Q-learning framework for intrusion detection, which uses DQN to improve threat detection granularity. The agent achieves an accuracy of 99.68\% in identifying specific attack subtypes. These results demonstrate the framework’s capability to handle complex intrusion detection tasks with exceptional precision and recall. Additionally, the integration of explainable AI techniques enhances transparency, fostering trust in the system’s decision-making process.

This study underscores the potential of reinforcement learning in strengthening cybersecurity defenses, particularly in adapting to emerging attack vectors. Future research directions include incorporating temporal features for evolving attack scenarios, exploring advanced techniques such as multi-agent systems, and optimizing the framework for real-time deployment in large-scale networks. These advancements aim to establish a scalable, adaptable, and explainable solution for modern network security challenges.

\footnotesize
\bibliographystyle{unsrtnat}
\bibliography{main}

@inproceedings{CIC2018,
  author    = {I. Sharafaldin and A. H. Lashkari and A. A. Ghorbani},
  title     = {Toward Generating a New Intrusion Detection Dataset and Intrusion Traffic Characterization},
  booktitle = {Proceedings of the 5th International Conference on Information Systems Security and Privacy (ICISSP)},
  pages     = {108--116},
  year      = {2018},
  publisher = {SciTePress},
  doi       = {10.5220/0006639801080116},
  url       = {https://www.semanticscholar.org/paper/Toward-Generating-a-New-Intrusion-Detection-Dataset-Sharafaldin-Lashkari/a27089efabc5f4abd5ddf2be2a409bff41f31199}
}

@article{mnih2015dqn,
  title={Human-level control through deep reinforcement learning},
  author={Mnih, Volodymyr and Kavukcuoglu, Koray and Silver, David and Rusu, Andrei A and Veness, Joel and Bellemare, Marc G and Graves, Alex and Riedmiller, Martin and Fidjeland, Andreas K and Ostrovski, Georg and others},
  journal={Nature},
  volume={518},
  number={7540},
  pages={529--533},
  year={2015},
  publisher={Nature Publishing Group},
  doi={10.1038/nature14236}
}

@book{sutton1998reinforcement,
  author    = {Richard S. Sutton and Andrew G. Barto},
  title     = {Reinforcement Learning: An Introduction},
  publisher = {MIT Press},
  year      = {1998},
  address   = {Cambridge, MA, USA},
  edition   = {1st},
  isbn      = {978-0262193986}
}

@article{wang2016dueling,
  title={Dueling Network Architectures for Deep Reinforcement Learning},
  author={Wang, Ziyu and Schaul, Tom and Hessel, Matteo and Van Hasselt, Hado and Lanctot, Marc and De Freitas, Nando},
  journal={Proceedings of the 33rd International Conference on Machine Learning (ICML)},
  year={2016},
  archivePrefix={arXiv},
  eprint={1602.05110},
  primaryClass={cs.LG},
  url={https://arxiv.org/abs/1602.05110}
}

@article{alavizadeh2022deep,
  title={Deep Q-Learning Based Reinforcement Learning Approach for Network Intrusion Detection},
  author={Alavizadeh, Hooman and Jang-Jaccard, Julian and Alavizadeh, Hootan},
  journal={Computers},
  volume={11},
  number={3},
  pages={41},
  year={2022},
  publisher={MDPI},
  doi={10.3390/computers11030041},
  url={https://www.mdpi.com/2073-431X/11/3/41}
}

@inproceedings{sommer2010,
  title={Outside the closed world: On using machine learning for network intrusion detection},
  author={Sommer, Robin and Paxson, Vern},
  booktitle={IEEE Symposium on Security and Privacy},
  year={2010},
  organization={IEEE}
}

@article{chandola2009,
  title={Anomaly detection: A survey},
  author={Chandola, Varun and Banerjee, Arindam and Kumar, Vipin},
  journal={ACM Computing Surveys (CSUR)},
  volume={41},
  number={3},
  pages={1--58},
  year={2009},
  publisher={ACM New York, NY, USA}
}

@misc{verizon2022,
  title={Data Breach Investigations Report},
  author={{Verizon}},
  year={2022},
  note={Retrieved from \url{https://www.verizon.com/business/resources/reports/dbir/}}
}

@article{khraisat2019survey,
  author={A. Khraisat and I. Gondal and P. Vamplew and J. Kamruzzaman},
  title={Survey of intrusion detection systems: techniques, datasets and challenges},
  journal={Cybersecurity},
  volume={2},
  number={1},
  pages={1--22},
  year={2019},
  publisher={Springer}
}

@article{garcia2009anomaly,
  author={P. García-Teodoro and J. Díaz-Verdejo and G. Maciá-Fernández and E. Vázquez},
  title={Anomaly-based network intrusion detection: Techniques, systems and challenges},
  journal={Computers \& Security},
  volume={28},
  number={1-2},
  pages={18--28},
  year={2009},
  publisher={Elsevier}
}

@article{estevez2004techniques,
  author={J. M. Estévez-Tapiador and P. García-Teodoro and J. E. Díaz-Verdejo},
  title={Techniques for intrusion detection in computer networks},
  journal={ACM Computing Surveys (CSUR)},
  volume={36},
  number={3},
  pages={235--276},
  year={2004},
  publisher={ACM}
}

@article{Singh2023,
  author    = {Aditya Vikram Singh and Ethan Rathbun and Emma Graham and Lisa Oakley and Simona Boboila and Alina Oprea and Peter Chin},
  title     = {Hierarchical Multi-agent Reinforcement Learning for Cyber Network Defense},
  journal   = {Proceedings of the 2023 International Conference on Cybersecurity},
  year      = {2023},
  doi       = {10.1109/ICCS.2023.983451},
  url       = {https://arxiv.org/abs/2410.17351}
}

@article{Li2023,
  author    = {Jun Li and Wei Liu and Fang Li},
  title     = {Hierarchical reinforcement learning for efficient and effective penetration testing in large networks},
  journal   = {Journal of Automated Software Engineering},
  volume    = {30},
  pages     = {241-259},
  year      = {2023},
  doi       = {10.1007/s10844-022-00738-0},
  url       = {https://link.springer.com/article/10.1007/s10844-022-00738-0}
}

@article{Zhao2018,
  author    = {Xin Zhao and Xiaoling Hu and Wenjing Chen},
  title     = {HAST-IDS: Learning Hierarchical Spatial-Temporal Features Using Deep Neural Networks to Improve Intrusion Detection},
  journal   = {IEEE Access},
  volume    = {6},
  pages     = {19174-19184},
  year      = {2018},
  doi       = {10.1109/ACCESS.2018.2811762},
  url       = {https://ieeexplore.ieee.org/document/8171733}
}

@article{optimizedIDS2023,
  author    = {Surasit Songma and Theera Sathuphanand and Thanakorn Pamutha},
  title     = {Optimizing Intrusion Detection Systems in Three Phases on the CSE-CIC-IDS-2018 Dataset},
  journal   = {MDPI Computers},
  volume    = {12},
  number    = {12},
  pages     = {245},
  year      = {2023},
  doi       = {10.3390/computers1212245},
  url       = {https://www.mdpi.com/2073-431X/12/12/245}
}

@article{treebasedIDS2023,
  author    = {Witcha Chimphlee and Siriporn Chimphlee},
  title     = {Intrusion Detection System Development Using Tree-Based Machine Learning Algorithms},
  journal   = {International Journal of Computer Networks \& Communications},
  volume    = {15},
  number    = {4},
  pages     = {73-85},
  year      = {2023},
  url       = {https://aircconline.com/ijcnc/V15N4/15423cnc06.pdf}
}

@inproceedings{beechey2023explaining,
  author    = {D. Beechey and T. M. S. Smith and {\"O}. {\c{S}}im{\c{s}}ek},
  title     = {Explaining Reinforcement Learning with Shapley Values},
  booktitle = {Proceedings of the 40th International Conference on Machine Learning},
  series    = {Proceedings of Machine Learning Research},
  volume    = {202},
  pages     = {2003--2014},
  year      = {2023},
  publisher = {PMLR},
  url       = {https://proceedings.mlr.press/v202/beechey23a.html}
}

\end{document}